 \documentclass[12pt]{article}

\usepackage{amsfonts}
\usepackage{amsmath}
\usepackage{amssymb}
\usepackage{amscd}
\usepackage{amsthm}
\usepackage{indentfirst}
\usepackage{epsfig}

\newtheorem{thm}{Theorem}[section]

\newtheorem{question}[thm]{Question}

\newcommand{\R}{{\mathbb{R}}}

\newcommand{\C}{{\mathbb{C}}}
\newcommand{\SP}{{\mathbb{S}}}

\newcommand{\cA}{{\mathcal{A}}}

\newcommand{\cM}{{\mathcal{M}}}

\newcommand{\cP}{{\mathcal{P}}}

\DeclareMathOperator{\osc}{osc}

\DeclareMathOperator{\tr}{tr}

\begin{document}

\title{An ``anti-Gleason'' phenomenon and simultaneous measurements in classical mechanics\\
}

\renewcommand{\thefootnote}{\alph{footnote}}

\author{\textsc Michael Entov$^{a}$,\ Leonid
Polterovich$^{b}$\ and Frol Zapolsky}

\footnotetext[1]{Partially supported by E. and J. Bishop Research
Fund and by the Israel Science Foundation grant $\#$ 881/06.} \footnotetext[2]{Partially supported by the Israel Science
Foundation grant $\#$ 11/03.}

\date{\today}

\maketitle

\begin{abstract}
\noindent We report on an ``anti-Gleason'' phenomenon in classical
mechanics: in contrast with the quantum case, the algebra of
classical observables can carry a non-linear quasi-state, a
monotone  functional which is linear on all subspaces generated by
Poisson-commutative functions. We present an example of such a
quasi-state in the case when the phase space is the 2-sphere. This
example lies in the intersection of two seemingly remote
mathematical theories -- symplectic topology and the theory of
topological quasi-states. We use this quasi-state to estimate the
error of the simultaneous measurement of non-commuting
Hamiltonians.
\end{abstract}



\renewcommand{\thefootnote}{\arabic{footnote}}

\section{Introduction}\label{intro}\noindent
Let $\cA_q$ (resp. $\cA_c$) be the algebra of observables in
quantum (resp. classical) mechanics. In the quantum case, $\cA_q$
is the space of hermitian operators on a Hilbert
space\footnote{For the sake of transparency, in this note we deal
with finite-dimensional Hilbert spaces only.} $H$.  It is equipped
with the bracket $[A,B]_{\hbar} = \frac{i}{\hbar}(AB-BA)\,,$ where
$\hbar$ is the Planck constant. In the classical case, $\cA_c$ is
the space of continuous real-valued functions on a symplectic
manifold $(M,\omega)$. It is equipped with the Poisson bracket
$\{\cdot,\cdot\}$ (defined on a dense subspace of smooth
functions). We say that observables $A$ and $B$ {\it commute} if
their bracket vanishes.

In \cite{von Neumann} von Neumann introduced the notion of a quantum
state. According to his definition, a \emph{state} is a functional
$\rho : \cA_q \to \R$ which satisfies:

\medskip
\noindent {\bf (Linearity)} $\rho(aA+bB) = a\rho(A)+b\rho(B)$ for
all $a,b \in \R$ and all $A,B \in \cA_q$;

\medskip
\noindent{\bf (Positivity)} $\rho(A)\geq 0$ provided $A \geq 0$;

\medskip
\noindent {\bf (Normalization)} $\rho(1) = 1$.
\medskip

This system of axioms implies that for every quantum state $\rho$
there exists {\it a density operator} $U$, that is a nonnegative
Hermitian operator  on \(H\) having trace \(1\), so that
$\rho(A)=\tr (UA)$ for every observable $A \in \cA_q$.

Von Neumann's  notion of quantum state encountered criticism among
physicists (for example, see \cite{Bell}) in that the formula
$\rho(A+B) = \rho(A) + \rho(B)$ \emph{a priori} makes sense only
if the observables $A$ and $B$ are simultaneously measurable,
which in the mathematical language means that they commute. As a
response to the criticism there appeared the concept of a
quasi-state. A \emph{quasi-state} on $\cA$ (here $\cA$ stands
either for $\cA_q$ or $\cA_c$) is a functional $\zeta : \cA \to
\R$ satisfying the positivity and normalization axioms above, and
a weaker form of linearity, the so-called

\medskip
\noindent {\bf (Quasi-linearity)} $\zeta(aA+bB) =
a\zeta(A)+b\zeta(B)$ for all $a,b \in \R$ and all \emph{commuting}
observables $A,B \in \cA$.
\medskip

A remarkable fact, which is a straightforward reformulation of the
famous theorem due to Gleason \cite{Gleason}, is as
follows\footnote{Gleason's paper does not mention quasi-states.
Bell \cite{Bell}, while analyzing Gleason's work, approaches this
notion without naming it. Another definition of a quasi-state,
equivalent to ours in the quantum setting, is given in \cite{Aar}.
See \cite{EP-qst} for a discussion on the link between the two
definitions in the classical setting.}:

\begin{thm}[Gleason]\label{thm-Gleason-prim}
On a Hilbert space of dimension at least $3$, every quasi-state is
linear.
\end{thm}

\noindent Therefore in the quantum case there is no distinction
between states and quasi-states.

Some brief comments on the Gleason theorem are in order. The key
notion responsible for Gleason's phenomenon and which is of an
independent significance in the subject of quantum logic is a
\emph{probability measure on projections}: Let \(\cP\) be the set
of orthogonal projection operators on \(H\), that is the simplest
observables which attain values \(1\) and \(0\) only. They can be
interpreted as yes-no experiments. Call two such operators \(P,Q
\in \cP\) \emph{orthogonal} to one another, if their ranges are
orthogonal as subspaces of \(H\). Note that orthogonal projections
commute. Given a  collection \(\{P_i\}\) of pairwise orthogonal
projections, their sum \(\sum_i P_i\) is again an orthogonal
projection. A probability measure on projections is a function
\(\mu : \cP \to [0,1]\) with \(\mu(1) = 1\) satisfying the
following {\it $\sigma$-additivity} property (cf.
\cite[p.132]{Jauch}): $\mu \big(\sum_{i} P_i \big) = \sum_{i} \mu
(P_i)$ for any collection $\{P_i\}$ of pairwise orthogonal
projection operators. By an ingenious argument Gleason shows
that\footnote{This is the main result of \cite{Gleason}.} for
every such $\mu$ there exists a density operator $U$ so that
\(\mu(P) = \tr (UP)\) for all \(P \in \cP\). Given any quasi-state
$\zeta$, its restriction to $\cP$ is a probability measure on
projections, and hence $\zeta(P) = \tr (UP)$ for some density
operator $U$. In view of the spectral decomposition, every
Hermitian operator $A \in \cA_q$ can be written as a real linear
combination of pairwise orthogonal projections. Thus
quasi-linearity of $\zeta$ yields $\zeta(A) = \tr (UA)$, and in
particular $\zeta$ is linear.

The Gleason theorem plays an important role in quantum mechanics.
It provides a counter-argument to the above-mentioned criticism of
von Neumann's concept of quantum state, and thus validates
representation of states by density operators which serve as a
cornerstone for probabilistic formalism in quantum mechanics.
Furthermore, the Gleason theorem stimulated a number of exciting
developments related to the hidden variables problem. For
instance, a starting point for the seminal works by Bell
\cite{Bell} and Kochen-Specker \cite{KS} is the following
observation: Existence of non-contextual hidden variables would
yield existence of dispersion-free states and hence supply the
quantum theory with the possibility of an assignment of a definite
outcome to every yes-no experiment. Such an assignment would be
nothing else but a probability measure on projections attaining
values $0$ and $1$ only. One readily sees that this is prohibited
by the Gleason theorem.

The purpose of this note is to show that in classical mechanics
the situation is quite different. In fact, we report on the
following "anti-Gleason phenomenon": {\it for certain symplectic
manifolds, the algebra of classical observables $\cA_c$ carries
non-linear quasi-states.} Since algebras of classical observables
often arise as a limit of algebras of quantum observables as
$\hbar \to 0$, our result can be interpreted as the failure of the
Gleason theorem in the classical limit. In Section~\ref{sec-quas}
we describe the simplest meaningful example of a non-linear
quasi-state in the case when the underlying symplectic manifold is
the $2$-sphere equipped with an area form. Such a phase space
appears for instance as the classical limit of the high spin
quantum particle, see Section~\ref{sec-spin}. As an application of
our quasi-state, we indicate in Section~\ref{sec-meas} that it
gives rise to a robust lower bound for the error of a simultaneous
measurement (in a sense to be made precise) of a pair of
non-commuting classical observables.

Complete formulations and proofs of the results discussed below can
be found in our papers \cite{EP-qmm,EP-qst,EPZ}.

\section{A non-linear quasi-state on $\SP^2$}\label{sec-quas}\noindent
Consider the unit $2$-sphere $\SP^2 \subset \R^3(x,y,z)$. Let
$\omega$ be the area form induced from the Euclidean metric and
divided by $4\pi$, so that the total area of the sphere equals
$1$.

\begin{figure}[h!t]
\begin{minipage}{6cm}
\epsfig{file=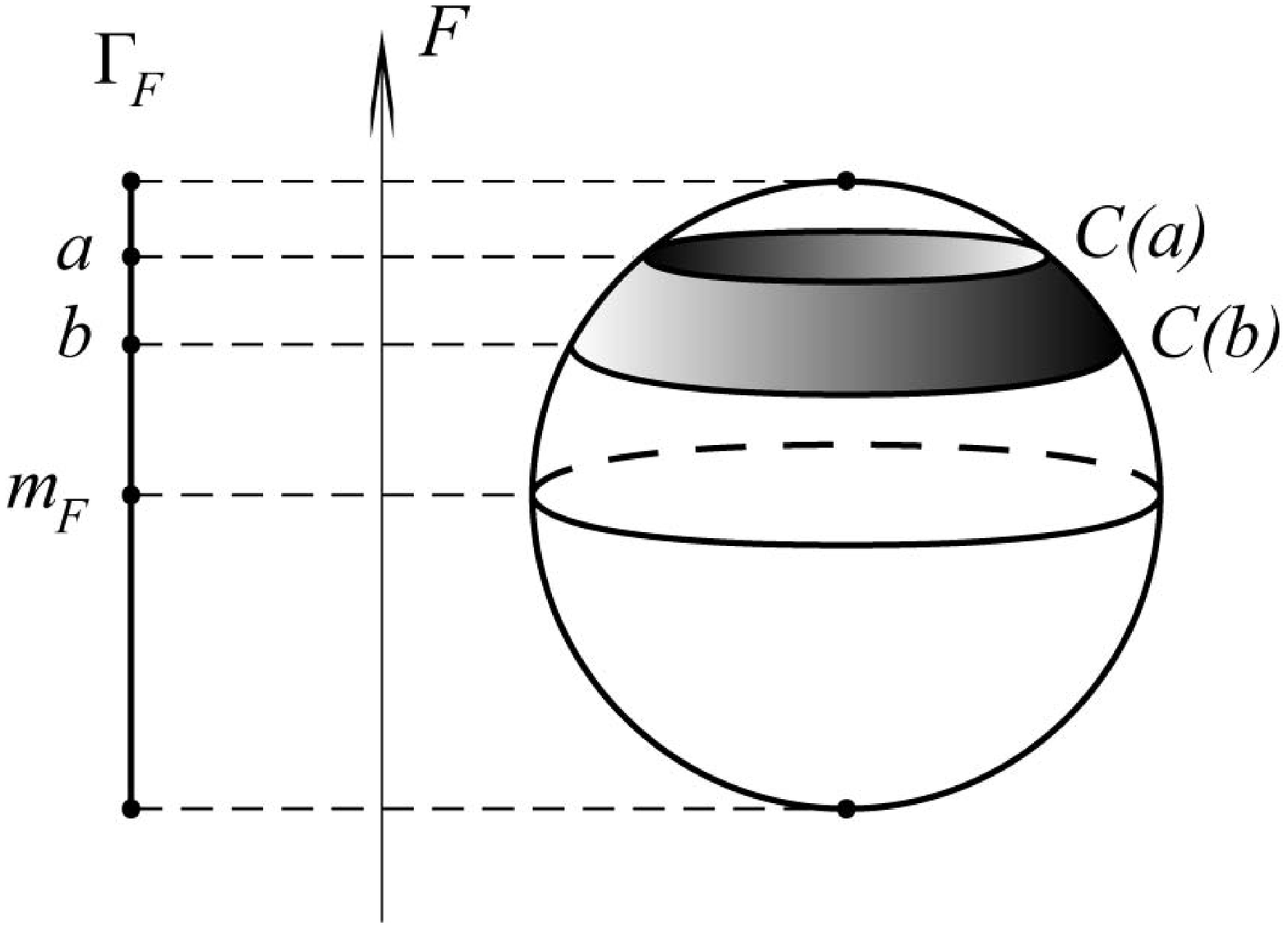,width=6cm} \caption{Round sphere. The
Reeb graph $\Gamma_F$ is a segment.} \label{dipode}
\end{minipage}
\hfill
\begin{minipage}{7.2cm}
\epsfig{file=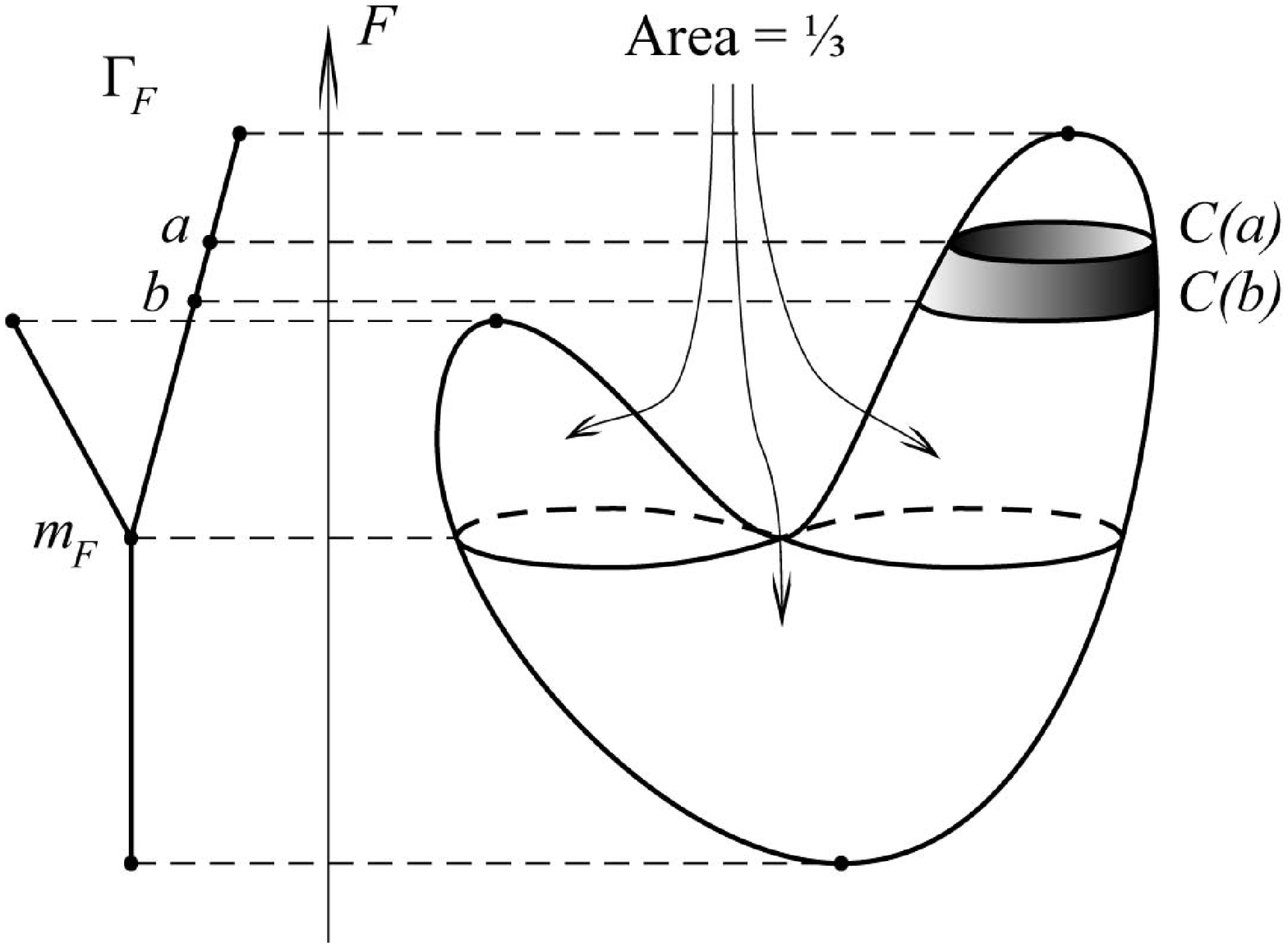,width=7.2cm} \caption{Deformed
sphere. The Reeb graph $\Gamma_F$ is a tripod.} \label{tripode}
\end{minipage}
\end{figure}

Let $F$ be a generic function\footnote{By a generic function we
mean a smooth function having only isolated critical points, whose
Hessian is non-degenerate at each such point, and whose critical
values are all distinct.} on $\SP^2$. Consider its {\it Reeb
graph} $\Gamma_F$ obtained from the sphere by collapsing each
connected component of each level set $\{F=\text{const}\}$ to a
point. This notion was first considered by Reeb \cite{Reeb} in the
framework of Morse theory (see e.g. \cite{Reeb1} for a detailed
discussion). It is illustrated on Figs.~\ref{dipode},
\ref{tripode}. Since it is hard to visualize a complicated
function on the round sphere $\SP^2$, we employ the following
trick: we represent the sphere as a surface in $\R^3$ (possibly of
complicated shape), and take $F$ to be the height function
$F(x,y,z) = z$ on it. It can be easily seen that $\Gamma_F$ is a
tree\footnote{A tree is a connected graph with no loops.}. For a
point $a \in \Gamma_F$ denote by $C(a)$ the corresponding
connected component of the level set. If $a$ is a vertex of the
graph, $C(a)$ is either a point of local extremum, or a ``figure
eight'' with the double point at a saddle. If $a$ is an interior
point of an edge, $C(a)$ is a simple closed curve on the sphere.

Introduce a probability measure $\sigma$ on $\Gamma_F$ in the
following way. Consider any open interval $I=(a,b)$ on an edge of
$\Gamma_F$. It corresponds to the annulus on $\SP^2$ bounded by
curves $C(a)$ and $C(b)$ (see Figs.~\ref{dipode}, \ref{tripode}).
By definition, the measure $\sigma(I)$ of the interval $I$ is
equal to the area of this annulus.

Given such a tree $\Gamma_F$ with the probability measure
$\sigma$, there exists a unique point $m_F$ on it, called {\it the
median} of the tree, with the following property: when $m_F$ is
removed from $\Gamma_F$, the resulting set breaks up into
connected components, each of measure $\leq \frac{1}{2}$  -- see
Figs.~\ref{dipode}, \ref{tripode}; on Fig.~\ref{dipode} the median
is the midpoint of the segment $\Gamma_F$; on Fig.~\ref{tripode}
the median is the triple point of the tripod.
Define $\zeta(F)$ as the value of $F$ on the level $C(m_F)$.

For example, if $F(x,y,z)=x$ (on the round $\SP^2$), the tree
$\Gamma_F$ is simply the segment $[-1;1]$, its median $m_F$ is the
point $0$, and the level $C(m_F)$ coincides with the equator
$\{x=0\}$. Hence
\begin{equation}\label{eq-x}\zeta(x) = 0\;.\end{equation}

We claim that $\zeta(F)\leq \zeta(G)$ for any pair of generic
functions $F$ and $G$ with $F \leq G$. Indeed, each connected
component of the sets $\SP^2 \setminus C(m_F)$ and $\SP^2
\setminus C(m_G)$ is a disc of area $\leq \frac{1}{2}$. Therefore
$C(m_F)$ and $C(m_G)$ must intersect at some point, say, $P$. Then
$$\zeta(F) = F(P)\leq G(P) = \zeta(G)$$ and the claim follows.

Write $\|F\|$ for the uniform norm $\max_{\SP^2}|F|$. The
monotonicity property above yields that $$|\zeta(F)-\zeta(G)|\leq
\|F-G\|$$ for all generic functions $F$ and $G$. Thus we can
extend $\zeta$ by continuity to the whole space $\cA_c$.

It turns out that $\zeta:\cA_c \to \R$ is a quasi-state.
Obviously, it satisfies the normalization and the positivity
axioms. Let us illustrate the quasi-linearity axiom. For
simplicity, we verify the property
$$\{F,G\}=0\Rightarrow \zeta(F+G)=\zeta(F)+\zeta(G)$$ in the
case when the functions $F,G$ and $F+G$ are generic. Note that in
this case the assumption $\{F,G\}=0$ simply means that $F,G$ and
$F+G$ are pairwise functionally dependent and therefore have the same
connected components of the level sets. One can easily conclude
that the curves $C(m_F),C(m_G)$ and $C(m_{F+G})$ coincide.
Denoting this curve by $C$ we have that for every point $P \in C$
$$\zeta(F+G)=(F+G)(P) = F(P) +G(P) =\zeta(F)+\zeta(G)\;,$$
as required.

We shall refer to the quasi-state $\zeta$ as {\it the median
quasi-state}.

Assume that $F:\SP^2 \to \R$ is a generic function. For a
continuous function $u:\R \to \R$ consider the composition
$G(x,y,z) = u(F(x,y,z))$, which can be already non-generic. It is
not hard to show that
\begin{equation}
\label{eq-compos} \zeta(G) = u(\zeta(F))\;.
\end{equation}
We shall apply this formula in order to calculate $\zeta(G)$, where
$G(x,y,z)=x^2$. Note that $G$ has a circle of minima points and
hence is not generic. Put $F(x,y,z) = x$ and $u(s) = s^2$ so that
$G(x,y,z) = u(F(x,y,z))$. Combining formulas \eqref{eq-compos} and
\eqref{eq-x} we get that $\zeta(x^2) = \zeta(x)^2= 0$. Similarly,
$\zeta(y^2)=\zeta(z^2)=0$.

Finally, let us verify that the median quasi-state is non-linear.
Indeed, since on $\SP^2$ we have $x^2+y^2+z^2 = 1$, it follows
from the definition of a quasi-state that $\zeta(x^2+y^2)=
1-\zeta(z^2)=1$. Thus
\begin{equation}\label{eq-nonlin}
\zeta(x^2+y^2)-\zeta(x^2)-\zeta(y^2) = 1\;.
\end{equation}

\section{Simultaneous classical measurements}\label{sec-meas}
\noindent Two non-commuting quantum observables are not
simultaneously measurable. Is there an analogous phenomenon in
classical mechanics? This problem appears in physics literature
(see e.g. books by Peres \cite[Chapter 12-2]{Peres} and Holland
\cite[Chapter 8.1]{Holland}) as a toy example motivating the
theory of quantum measurements. Theoretically, in a classical
system any two observables are simultaneously measurable to any
accuracy. However, if the measurement is not perfect, an error may
appear. Below we present a precise formulation of these heuristic
notions and give a positive answer to the above question.


We shall analyze simultaneous measurability in classical mechanics
in the framework of a measurement procedure called the {\it
pointer model}. For simplicity we work on the sphere $\SP^2$ and
write $\zeta$ for the median quasi-state introduced in the
previous section. We denote by $\|F\|$ the uniform norm of a
function $F$ on $\SP^2$ and by $\langle F \rangle$ its mean value
$\int_{\SP^2} F \cdot \omega$. For a pair of smooth functions
$F_1,F_2$ on the sphere consider the quantity
$$\Pi(F_1,F_2)= |\zeta(F_1+F_2)-\zeta(F_1)-\zeta(F_2)|$$
which measures the non-additivity of $\zeta$ at this pair. Define
also the oscillation
$$\osc(F_1,F_2) = \min \big( \| F_1
-\langle F_1 \rangle \|,\| F_2 -\langle F_2 \rangle \| \big)\;.$$

Consider two observables $F_1,F_2 \in \cA_c$. Let $ M = \SP^2
\times \R^{4}(p_1, q_1, p_2, q_2)$ be the extended phase space
equipped with the symplectic form\footnote{For preliminaries on
symplectic geometry see, for example, \cite{Berndt}.} $\widehat
\omega = \omega + dp_1\wedge dq_1 + dp_2\wedge dq_2$ . The $\R^4$
factor corresponds to the measuring apparatus (the pointer),
whereas $\vec q = (q_1,q_2)$ is the quantity read from it. The
coupling of the apparatus to the system is carried out by means of
the Hamiltonian function $p_1F_1(v)+p_2F_2(v)$, $v \in \SP^2$. The
Hamiltonian equations of motion with the initial conditions
$q_1(0)=q_2(0)=0, p_1(0)=p_2(0) = \epsilon$ and $v(0) = w$ are as
follows:
\begin{align*}
\dot {q_i} &= F_i,\; i=1,2,\\
\dot {p_i} &= 0,\; i = 1,2,\\
\dot v &= \epsilon\ V_{F_1+F_2}(v)\;.
\end{align*}
Here $V_{F_1 + F_2}$ denotes the Hamiltonian vector field of the
Hamiltonian $F_1 + F_2$ on $\SP^2$.

Denote by $g_t$ the Hamiltonian flow on $\SP^2$ generated by the
function $G=F_1+F_2$. Then $v(t) = g_{\epsilon t}w$. Let $T>0$ be
the duration of the measurement. By definition, the output of the
measurement procedure is a pair of functions $F'_i, \; i=1,2,$ on
$M$ defined by the average displacement of the $q_i$-coordinate of
the pointer:
$$F'_i (w) = \frac{1}{T}(q_i(T) -q_i(0)) =\frac{1}{T}\int_0^T
F_i(v(t))dt =\frac{1}{T}\int_0^T F_i(g_{\epsilon t}w)dt \;.$$ Note
that for $\epsilon =0$ we have $F'_i =F_i$. This justifies the above
procedure as a measurement of $F_i$ and allows us to interpret the
number $\epsilon$ as {\it an imprecision of the pointer}.

Define {\it the error of the measurement} as
\[\Delta(T,\epsilon,F_1,F_2) = \left\| F'_i - F_i \right\| .\]
Note that in our setting this quantity does not depend on $i \in
\{1;2\}$ since the sum $F_1+F_2$ is constant along the
trajectories of $g_t$.

Now we are ready to formulate our main result \cite{EPZ}: for all
$T,\epsilon
> 0$ and $F_1,F_2 \in \cA_c$
\begin{equation}\label{thm-meas}\Delta(T,\epsilon,F_1,F_2) \geq
\frac{1}{2}\Pi(F_1,F_2)-\sqrt{\frac{2\cdot \osc(F_1,F_2)
}{T\epsilon}}\;.
\end{equation}

For $\epsilon>0$ define the asymptotic (as $T \to \infty$) error
of the measurement as
$$\Delta_{\infty}(F_1,F_2) = \liminf_{T\to
\infty}\Delta(T,\epsilon,F_1,F_2)\;.$$ It may be interpreted as
the error produced in a system moving very rapidly, that is such
that its characteristic time is much less than that of a
measurement. Note that $\Delta(T,\epsilon,F_1,F_2) =
\Delta(\epsilon T, 1, F_1, F_2)$ and hence $\Delta_{\infty}$ does
not depend on the specific choice of $\epsilon >0$. It follows
from inequality \eqref{thm-meas} that
\begin{equation}\label{eq-asymp}
\Delta_{\infty}(F_1,F_2) \geq \frac{1}{2}\Pi(F_1,F_2)\;.
\end{equation}

Define a numerical constant
$$E := \inf_{F_1,F_2} \frac{2\Delta_{\infty}(F_1,F_2)}{\Pi(F_1,F_2)}\;,$$ where the infimum
is taken over all pairs of smooth observables $F_1$ and $F_2$ on
the sphere with $\Pi(F_1,F_2) \neq 0$. Note that $E \geq 1$ in
view of $\eqref{eq-asymp}$. To find an upper bound on $E$,
consider the case $F_1 = x^2$ and $F_2 = y^2$. An elementary but
cumbersome calculation shows that $\Delta_{\infty}(x^2,y^2) \sim
0.63$. Equation \eqref{eq-nonlin} above yields $\Pi(x^2,y^2) = 1$.
Therefore $E \leq 1.26$. It would be interesting to calculate the
value of $E$ explicitly.

Let us emphasize a somewhat surprising feature of inequality
\eqref{thm-meas}. Its right-hand side is robust with respect to
small perturbations of both observables in the uniform norm. On
the other hand, the measurement error $\Delta$ involves the
Hamiltonian flow generated by $F_1+F_2$ which is defined by {\it
the first derivatives} of $F_1$ and $F_2$. Therefore {\it a
priori} $\Delta(T,\epsilon,F_1,F_2)$ could have changed in an
arbitrary way after such a perturbation, in particular it could
have vanished, but this does not happen provided $\Pi(F_1,F_2)
\neq 0$.

It is instructive to mention that our lower bound \eqref{thm-meas}
on the error of the simultaneous measurement of a pair of
classical non-commuting observables $F_1,F_2$ with
$\Pi(F_1,F_2)\neq 0$ cannot be considered as a classical version
of the uncertainty principle. Indeed, the quantum uncertainty
principle deals with the statistical dispersion of similarly
prepared systems \cite[p.379]{Ballentine}. Let us interpret for a
moment the quantity $\zeta(F)$ as the statistical expectation of
the value of the observable $F$ in the (quasi-)state $\zeta$. With
this language, the quasi-state $\zeta$ introduced in
Section~\ref{sec-quas} is {\it dispersion free:}
$\zeta(F^2)-\zeta(F)^2 =0$ for all $F$ (see \cite{EP-qst}).

\section{Discussion: the classical limit of quantum spin system
}\label{sec-spin}\noindent It is an intriguing question \cite{H}
to understand ``what went wrong'' with the proof of the Gleason
theorem in the classical context. Our impression is that the
notion of the probability measures on projections, which is the
key character of the proof (see Section~\ref{intro} above), does
not admit a natural translation into the classical language via
the quantum-classical correspondence. However, instead of
considering the ``anti-Gleason'' phenomenon as a failure of the
correspondence principle, we propose to address the above question
the other way around:

\medskip
\noindent
\begin{question}\label{quest-foot}
Is there a footprint of the median quasi-state $\zeta$ in the
quantum world?
\end{question}

\medskip
\noindent To be more specific, recall that the algebra of
functions on the 2-sphere $\SP^2$ equipped with the Poisson
bracket arises as the classical limit of a quantum spin system.
Below we briefly review a construction called {\it the coherent
states (de)quantization} \cite{Be, Pe}. Consider a quantum
particle of spin $j$ where $j>0$ is an integer or a half integer
number. The quantum states are modeled by vectors of a Hilbert
space $H$ of dimension $N=2j+1$ which carries an irreducible
representation of the group $SU(2)$. Denote by $K:= S^1 \subset
SU(2)$ the subgroup consisting of diagonal matrices. A standard
argument of the representation theory yields the existence of a
$K$-invariant complex line $\ell \subset H$ such that $K$ acts on
$\ell$ as the full $2j$-turn rotation. Fix a unit vector
$|\psi\rangle \in \ell$.

Denote by $\cM(N)$ the algebra of complex linear operators of $H$.
Given an operator $A \in \cM(N)$, define a function $\hat{s}_A:
SU(2) \to \C$ by $\hat{s}_A(g)= \langle g\psi|Ag \psi\rangle$. A
crucial observation is as follows: given $k \in K$, we have $k
|\psi\rangle= e^{i\theta} |\psi\rangle$ for some $\theta \in \R$,
and therefore $\hat{s}_A(gk)=\hat{s}_A(g)$ for all $g \in SU(2), k
\in K$. This means that $\hat{s}_A$ descends to a smooth function,
say $s_A$, on the quotient space $SU(2)/K$. This quotient space is
naturally identified with the 2-sphere $\SP^2$. The function $s_A$
on $\SP^2$ is called the {\it covariant symbol} of the operator
$A$. Note that the correspondence
$$\cM(N) \to C^{\infty}(\SP^2), \; A \mapsto s_A$$ sends
the space of Hermitian operators $\cA_q(N) \subset \cM(N)$ to the
space $\cA_c$ of real-valued functions, and non-negative operators
to non-negative functions.

Given any smooth function $a$ on $\SP^2$, there exists a sequence
of operators $A(N) \in \cM(N)$ so that the corresponding covariant
symbols $s_{A(N)}$ converge to $a$ as $N \to \infty$. Suppose now
that $s_{A(N)} \to a$ and $s_{B(N)} \to b$ for a pair of functions
$a,b $ on $\SP^2$. It turns out that the following correspondence
principle holds true:\footnote{In the model discussed below the
Planck constant $\hbar$ equals $\frac{1}{2j}$. Thus the classical
limit $\hbar \to 0$ is the high spin limit $N \to \infty$.}
$$s_{A_NB_N} \to ab,\;\; \text{and}\;\; N\cdot s_{[A(N),B(N)]} \to
i\{a,b\}\;\;\text{as}\;\;N \to \infty.$$ This can be interpreted
as follows: the algebra of classical observables $\cA_c$ on the
2-sphere $\SP^2$ equipped with the Poisson bracket arises as the
high spin limit $N \to \infty$  of algebras $\cA_q(N)$ of quantum
observables. Let us also mention that a related construction of
the {\it fuzzy sphere} \cite{Madore} paved a road for extension of
geometric analysis on the classical phase space into the framework
of matrix algebras.

In light of the discussion above, we can address
Question~\ref{quest-foot} in a slightly more precise way. The
Gleason theorem rules out the existence of a non-linear
quasi-state on $\cA_q(N)$ for every given value of $N$. Do the
algebras $\cA_q$ carry a weaker object (a kind of "approximate
quasi-state" still to be defined) which converges to the median
quasi-state $\zeta$ in the classical limit?

\section{Conclusion}\noindent
We have discussed the simplest version of the ``anti-Gleason''
phenomenon in classical mechanics by presenting the median
quasi-state on the algebra of classical observables of the
2-sphere. This result can be generalized in two directions.

First, quasi-states do exist on certain higher-dimensional
symplectic manifolds such as products $\SP^2 \times \dots \times
\SP^2$ and complex projective spaces $\C P^n$ (see \cite{EP-qst}).
They can be detected by methods of modern symplectic topology,
most notably by the Gromov-Floer theory (see e.g. \cite{MS}).
Second, closed 2-dimensional symplectic manifolds (i.e. closed
oriented surfaces equipped with an area form) carry a lot of
quasi-states of quite a different nature. They are provided by the
theory of {\it topological quasi-states} developed by Aarnes and
his collaborators -- see e.g. \cite{Aar,Q-function}. An
interesting feature of the median quasi-state $\zeta$ is that it
lies in the intersection of both (seemingly remote!)
above-mentioned areas of mathematics. Analyzing historical origins
of the notion of a quantum-mechanical quasi-state, it is tempting
to interpret dispersion-free quasi-states on the algebra of
classical observables as "hidden variables in classical
mechanics". It will become clearer after further exploration
whether such an interpretation is justified. In particular, it
would be interesting to explore dynamical features of the
evolution on the space of topological quasi-states induced by a
Hamiltonian flow $g_t$ on the underlying surface. This evolution,
say $L_t$, is defined via the Koopman operator associated to the
flow: Given a quasi-state $\rho$,  we set $L_t\rho(F) = \rho(F
\circ g_t)$. Let us emphasize that the median quasi-state on the
sphere is invariant under {\it any} Hamiltonian flow, and in this
respect it is an analogue of the Liouville measure considered as a
classical state.

Furthermore, we presented a lower bound \eqref{thm-meas} on the
error of the simultaneous measurement of a pair of non-commuting
classical observables. The bound is given in terms of the median
quasi-state and hence is robust with respect to small perturbations
of observables in the uniform norm. Similar bounds exist for certain
higher-dimensional symplectic manifolds. There is a good chance that
they can be also extended to dispersion-free topological
quasi-states on surfaces by using methods of a recent work \cite{Z}.

Let us emphasize that the error of a simultaneous measurement
arising in our model is caused by a combination of two factors:
non-commutativity of the observables and imprecision of the
measuring apparatus. P.~Busch \cite{Busch} brought our attention
to the work \cite{BB} which establishes a phenomenon of a quantum
mechanical nature, a violation of the Bell inequality, in the
context of fuzzy classical observables. It would be interesting to
explore whether an analogous phenomenon holds true within the
formalism of quasi-states. This leads to an intriguing
mathematical problem of finding restrictions on quasi-states for
composite classical systems, or, in other words, on cartesian
products of several symplectic manifolds.

\bigskip

\noindent{\bf Acknowledgments.}  We are grateful to Larry Horwitz
for very useful comments on the first draft of the paper and
stimulating questions. We thank Paul Busch and George Zaslavsky for
interesting discussions. We thank the anonymous referee for drawing
our attention to the high spin example and Constantin Brif and Ady
Mann for a valuable consultation on this subject.

\bibliographystyle{alpha}

\begin{thebibliography}{99}

\bibitem{Aar} Aarnes, J.F.,
{\it Quasi-states and quasi-measures, } Adv. Math. {\bf 86}:1
(1991), 41-67.

\bibitem{Q-function} Aarnes, J.F., Rustad, A.B. {\it Probability
and quasi-measures -- a new interpretation}, Math. Scand. {\bf 85}
(1999), 278--284.

\bibitem{Ballentine} Ballentine, L.E.,
{\it The statistical interpretation of quantum mechanics, } Rev.
Mod. Phys. {\bf 42} (1970), 358-381.

\bibitem{Bell} Bell, J.S., {\it On the problem of hidden variables in
quantum mechanics, } Rev. Modern Phys. {\bf 38} (1966), 447-452.

\bibitem{BB} Beltrametti, E.G., Bugajski, S.,
{\it The Bell phenomenon in classical frameworks,} J. Phys. A {\bf
29} (1996), 247-261.

\bibitem{Be} Berezin, F.A., {\it General concept of quantization,}
Comm. Math. Phys. {\bf 40} (1975), 153-174.


\bibitem{Berndt} Berndt, R., {\it An introduction to symplectic geometry},
Graduate Studies in Mathematics, {\bf 26}. American Mathematical
Society, 2000.

\bibitem{Busch} Busch, P., private communication.

\bibitem{Reeb1} Cole-McLaughlin, K., Edelsbrunner, H.,  Harer, J., Natarajan, V., Pascucci, V.,
{\it Loops in Reeb graphs of 2-manifolds}, Discrete Comput. Geom.
{\bf 32} (2004), 231-244.

\bibitem{EP-qmm} Entov, M., Polterovich, L., {\it Calabi quasimorphism
and quantum homology, } Intern. Math. Res. Notices {\bf 30}
(2003), 1635-1676.

\bibitem{EP-qst} Entov, M., Polterovich, L.,
{\it Quasi-states and symplectic intersections, } Comm. Math.
Helv. {\bf 81}:1 (2006), 75-99.

\bibitem{EPZ} Entov, M., Polterovich, L., Zapolsky, F.,
{\it Quasi-morphisms and the Poisson bracket}, preprint, arXiv
math.SG/0605406, 2006 (to appear in Pure and Applied Mathematics
Quaterly, a special issue in honor of G.~Margulis).

\bibitem{Gleason} Gleason, A.M.,
{\it Measures on the closed subspaces of a Hilbert space,} J.
Math. Mech. {\bf 6} (1957), 885-893.


\bibitem{Holland} Holland, P.R., {\it The quantum theory of motion. An account of
the de Broglie-Bohm causal interpretation of quantum mechanics.}
Cambridge University Press, Cambridge, 1995.


\bibitem{H} Horwitz, L., private communication.

\bibitem{Jauch} Jauch, J.M., {\it  Foundations of quantum mechanics.}
Addison-Wesley Publishing Co., Reading, Mass.-London-Don Mills,
Ont. 1968.

\bibitem{KS} Kochen, S., Specker, E.,
{\it The problem of hidden variables in quantum mechanics,} J.
Math. Mech. {\bf 17} (1967), 59-87.


\bibitem{Madore} Madore, J.,
{\it The fuzzy sphere, } Classical and Quantum Gravity {\bf 9}
(1992), 69-87.

\bibitem{MS} McDuff, D., Salamon, D., {\it
$J$-holomorphic curves and symplectic topology,} AMS, Providence,
2004.

\bibitem{Pe} Perelomov, A.,
{\it Generalized coherent states and their applications,} Texts and
Monographs in Physics. Springer-Verlag, Berlin, 1986.

\bibitem{Peres} Peres, A., {\it Quantum theory: concepts and
methods.} Fundamental Theories of Physics, 57. Kluwer Academic
Publishers Group, Dordrecht, 1993.

\bibitem{Reeb} Reeb, G.,
{\it Sur les points singuliers d'une forme de Pfaff completement
integrable ou d'une fonction numerique.} C. R. Acad. Sci. Paris
{\bf 222} (1946), 847--849.

\bibitem{von Neumann} von Neumann, J., {\it Mathematical foundations of
quantum mechanics}, Princeton University Press, Princeton, 1955.
(Translation of {\it Mathematische Grundlagen der
Quantenmechanik}, Springer, Berlin, 1932.)


\bibitem{Z} Zapolsky, F., {\it Quasi-states and the Poisson
bracket on surfaces}, preprint, arXiv:math/0703121 (to appear in
Journal of Modern Dynamics).


\end{thebibliography}

\bigskip

\noindent
\begin{tabular}{l @{\ \ \ \ \ \ \ \ \ \ \,} l }
Michael Entov & Leonid Polterovich \\
Department of Mathematics & School of Mathematical Sciences \\
Technion & Tel Aviv University \\
Haifa 32000, Israel & Tel Aviv 69978, Israel \\
entov@math.technion.ac.il &
polterov@post.tau.ac.il\\
\end{tabular}

\bigskip
\noindent
\begin{tabular}{l}
Frol Zapolsky \\
School of Mathematical Sciences \\
Tel Aviv University \\
Tel Aviv 69978, Israel \\
zapolsky@post.tau.ac.il \\
\end{tabular}

\end{document}